# NEW MULTICATEGORY BOOSTING ALGORITHMS BASED ON MULTICATEGORY FISHER-CONSISTENT LOSSES

BY HUI ZOU[1], JI ZHU AND TREVOR HASTIE

*University of Minnesota, University of Michigan and Stanford University*

Fisher-consistent loss functions play a fundamental role in the construction of successful binary margin-based classifiers. In this paper we establish the Fisher-consistency condition for multicategory classification problems. Our approach uses the margin vector concept which can be regarded as a multicategory generalization of the binary margin. We characterize a wide class of smooth convex loss functions that are Fisher-consistent for multicategory classification. We then consider using the margin-vector-based loss functions to derive multicategory boosting algorithms. In particular, we derive two new multicategory boosting algorithms by using the exponential and logistic regression losses.

**1. Introduction.** The margin-based classifiers, including the support vector machine (SVM) [Vapnik (1996)] and boosting [Freund and Schapire (1997)], have demonstrated their excellent performances in binary classification problems. Recent statistical theory regards binary margin-based classifiers as regularized empirical risk minimizers with proper loss functions. Friedman, Hastie and Tibshirani (2000) showed that AdaBoost minimizes the novel exponential loss by fitting a forward stage-wise additive model. In the same spirit, Lin (2002) showed that the SVM solves a penalized hinge loss problem and the population minimizer of the hinge loss is exactly the Bayes rule, thus, the SVM directly approximates the Bayes rule without estimating the conditional class probability. Furthermore, Lin (2004) introduced the concept of Fisher-consistent loss in binary classification and he showed that any Fisher-consistent loss can be used to construct a binary margin-based classifier. Buja, Stuetzle and Shen (2005) discussed the proper scoring rules for binary classification and probability estimation which are closely related to the Fisher-consistent losses.

Received March 2008; revised August 12.
[1]Supported by NSF grant DMS 07-06-733.

In the binary classification case, the Fisher-consistent loss function theory is often used to help us understand the successes of some margin-based classifiers, for the popular classifiers were proposed before the loss function theory. However, the important result in Lin (2004) suggests that it is possible to go the other direction: we can first design a nice Fisher-consistent loss function and then derive the corresponding margin-based classifier. This viewpoint is particularly beneficial in the case of multicategory classification. There has been a considerable amount of work in the literature to extend the binary margin-based classifiers to the multi-category case. A widely used strategy for solving the multi-category classification problem is to employ the *one-versus-all* method [Allwein, Schapire and Singer (2000)], such that a $m$-class problem is reduced to $m$ binary classification problems. Rifkin and Klautau (2004) gave very provocative arguments to support the one-versus-all method. AdaBoost.MH [Schapire and Singer (1999)] is a successful example of the one-versus-all approach which solves a $m$-class problem by applying AdaBoost to $m$ binary classification problems. However, the one-versus-all approach could perform poorly with the SVM if there is no dominating class, as shown by Lee, Lin and Wahba (2004). To fix this problem, Lee, Lin and Wahba (2004) proposed the multicategory SVM. Their approach was further analyzed in Zhang (2004a). Liu and Shen (2006) and Liu, Shen and Doss (2005) proposed the multicategory *psi*-machine.

In this paper we extend Lin's Fisher-consistency result to multicategory classification problems. We define the Fisher-consistent loss in the context of multicategory classification. Our approach is based on the margin vector, which is the multicategory generalization of the margin in binary classification. We then characterize a family of convex losses which are Fisher-consistent. With a multicategory Fisher-consistent loss function, one can produce a multicategory boosting algorithm by employing gradient decent to minimize the empirical margin-vector-based loss. To demonstrate this idea, we derive two new multicategory boosting algorithms.

The rest of the paper is organized as follows. In Section 2 we briefly review binary margin-based classifiers. Section 3 contains the definition of multicategory Fisher-consistent losses. In Section 4 we characterize a class of convex multicategory Fisher-consistent losses. In Section 5 we introduce two new multicategory boosting algorithms that are tested on benchmark data sets. Technical proofs are relegated to the Appendix.

**2. Review of binary margin-based losses and classifiers.** In standard classification problems we want to predict the label using a set of features. $y \in \mathcal{C}$ is the label where $\mathcal{C}$ is a discrete set of size $m$, and $\mathbf{x}$ denotes the feature vector. A classification rule $\delta$ is a mapping from $\mathbf{x}$ to $\mathcal{C}$ such that a label $\delta(\mathbf{x})$ is assigned to the data point $\mathbf{x}$. Under the 0–1 loss, the misclassification error of $\delta$ is $R(\delta) = P(y \neq \delta(\mathbf{x}))$. The smallest classification error is achieved by the Bayes rule $\arg\max_{c_i \in \mathcal{C}} p(y = c_i|\mathbf{x})$. The conditional class probabilities $p(y = c_i|\mathbf{x})$ are unknown, so is the Bayes rule. One must construct a classifier $\delta$ based on $n$ training samples $(y_i, \mathbf{x}_i), i = 1, 2, \ldots, n$, which are independent identically distributed (i.i.d.) samples from the underlying joint distribution $p(y, \mathbf{x})$.

In the book by Hastie, Tibshirani and Friedman (2001) readers can find detailed explanations of the support vector machine and boosting. Here we briefly discuss a unified statistical view of the binary margin-based classifier. In the binary classification problem, $\mathcal{C}$ is conveniently coded as $\{1, -1\}$, which is important for the binary margin-based classifiers. Consider a margin-based loss function $\phi(y, f) = \phi(yf)$, where the quantity $yf$ is called the margin. We define the empirical $\phi$ risk as $\text{EMR}_n(\phi, f) = \frac{1}{n}\sum_{i=1}^n \phi(y_i f(\mathbf{x}_i))$. Then a binary margin-based $\phi$ classifier is obtained by solving

$$\hat{f}^{(n)} = \arg\min_{f \in \mathcal{F}_n} \text{EMR}_n(\phi, f),$$

where $\mathcal{F}_n$ denotes a regularized functional space. The margin-based classifier is $sign(\hat{f}^{(n)}(\mathbf{x}))$. For the SVM, $\phi$ is the hinge loss and $\mathcal{F}_n$ is the collection of penalized kernel estimators. AdaBoost amounts to using the exponential loss $\phi(y, f) = \exp(-yf)$ and $\mathcal{F}_n$ is the space of decision trees. The loss function plays a fundamental role in the margin-based classification. Friedman, Hastie and Tibshirani (2000) justified AdaBoost by showing that the population minimizer of the exponential loss is one-half the log-odds. Similarly, in the SVM case, Lin (2002) proved that the population minimizer of the hinge loss is exactly the Bayes rule.

Lin (2004) further discussed a class of Fisher-consistent losses. A loss function $\phi$ is said to be Fisher-consistent if

$$\hat{f}(\mathbf{x}) = \arg\min_{f(\mathbf{x})}[\phi(f(\mathbf{x}))p(y = 1|\mathbf{x}) + \phi(-f(\mathbf{x}))p(y = -1|\mathbf{x})]$$

has a unique solution $\hat{f}(\mathbf{x})$ and

$$sign(\hat{f}(\mathbf{x})) = sign(p(y = 1|\mathbf{x}) - 1/2).$$

The Fisher-consistent condition basically says that with infinite samples, one can exactly recover the Bayes rule by minimizing the $\phi$ loss.

**3. Multicategory Fisher-consistent losses.** In this section we extend Lin's Fisher-consistent loss idea to the multicategory case. We let $\mathcal{C} = \{1, 2, \ldots, m\}$ ($m \geq 3$). From the definition of the binary Fisher-consistent loss, we can regard the margin as an effective proxy for the conditional class probability, if the decision boundary implied by the "optimal" margin is identical to the Bayes decision boundary. To better illustrate this interpretation of the margin, recall that $sign(p(y=1|\mathbf{x}) - 1/2)$ is the Bayes rule for binary classification and

$$sign(p(y=1|\mathbf{x}) - 1/2) = sign(p(y=1|\mathbf{x}) - p(y=-1|\mathbf{x})),$$
$$sign(\hat{f}(\mathbf{x})) = sign(\hat{f}(\mathbf{x}) - (-\hat{f}(\mathbf{x}))).$$

The binary margin is defined as $yf$. Since $yf = f$ or $-f$, an equivalent formulation is to assign margin $f$ to class 1 and margin $-f$ to class $-1$. We regard $f$ as the proxy of $p(y=1|\mathbf{x})$ and $-f$ as the proxy of $p(y=-1|\mathbf{x})$, for the purpose of comparison. Then the Fisher-consistent loss is nothing but an effective device to produce the margins that are a legitimate proxy of the conditional class probabilities, in the sense that the class with the largest conditional probability always has the largest margin.

We show that the proxy interpretation of the margin offers a graceful multicategory generalization of the margin. The multicategory margin is conceptually identical to the binary margin, which we call the margin-vector. We define the margin vector together with the multicategory Fisher-consistent loss function.

DEFINITION 1. A $m$-vector $f$ is said to be a *margin vector* if

$$\sum_{j=1}^{m} f_j = 0. \tag{3.1}$$

Suppose $\phi(\cdot)$ is a loss function and $f(\mathbf{x})$ is a margin vector for all $\mathbf{x}$. Let $p_j = p(y=j|\mathbf{x})$, $j = 1, 2, \ldots, m$, be the conditional class probabilities and denote $p = (\cdots p_j \cdots)$. Then we define the expected $\phi$ risk at $\mathbf{x}$:

$$\phi(p, f(\mathbf{x})) = \sum_{j=1}^{m} \phi(f_j(\mathbf{x})) p(y=j|\mathbf{x}). \tag{3.2}$$

Given $n$ i.i.d. samples, the empirical margin-vector based $\phi$ risk is given by

$$\text{EMR}_n(\phi) = \frac{1}{n} \sum_{i=1}^{n} \phi(f_{y_i}(\mathbf{x}_i)). \tag{3.3}$$

A loss function $\phi(\cdot)$ is said to be *Fisher-consistent* for $m$-class classification if $\forall \mathbf{x}$ in a set of full measure, the following optimization problem

$$\hat{f}(\mathbf{x}) = \arg\min_{f(\mathbf{x})} \phi(p, f(\mathbf{x})) \quad \text{subject to} \quad \sum_{j=1}^{m} f_j(\mathbf{x}) = 0 \tag{3.4}$$

has a unique solution $\hat{f}$, and

$$\arg\max_{j} \hat{f}_j(\mathbf{x}) = \arg\max_{j} p(y=j|\mathbf{x}). \tag{3.5}$$

Furthermore, a loss function $\phi$ is said to be *universally Fisher-consistent* if $\phi$ is Fisher-consistent for $m$-class classification $\forall m \geq 2$.

We have several remarks.

REMARK 1. We assign a margin $f_j$ to class $j$ as the proxy of the conditional class probability $p(y = j|\mathbf{x})$. The margin vector satisfies the sum-to-zero constraint such that when $m = 2$, the margin vector becomes the usual binary margin. The sum-to-zero constraint also ensures the existence and uniqueness of the solution to (3.3). The sum-to-zero constraint was also used in Lee, Lin and Wahba (2004).

REMARK 2. We do not need any special coding scheme for $y$ in our approach, which is very different from the proposal in Lee, Lin and Wahba (2004). The data point $(y_i, \mathbf{x}_i)$ belongs to class $y_i$, hence, its margin is $f_{y_i}(\mathbf{x}_i)$ and its margin-based risk is $\phi(f_{y_i}(\mathbf{x}_i))$. Thus, the empirical risk is defined as that in (3.3). If we only know $\mathbf{x}$, then $y$ can be any class $j$ with probability $p(y = j|\mathbf{x})$, hence, we consider the expected risk defined in (3.2).

REMARK 3. The Fisher-consistent condition is a direct generalization of the definition of the Fisher-consistent loss in binary classification. It serves the same purpose: to produce a margin vector that is a legitimate proxy of the conditional class probabilities such that comparing the margins leads to the multicategory Bayes rule.

REMARK 4. There are many nice Fisher-consistent loss functions for binary classification. It would be interesting to check if these losses for binary classification are also Fisher-consistent for multicategory problems. This question will be investigated in Section 4 where we show that most of popular loss functions for binary classification are universally Fisher-consistent.

REMARK 5. Buja, Stuetzle and Shen (2005) showed the connection between Fisher-consistent losses and *proper scoring rules* which estimate the class probabilities in a Fisher consistent manner. Of course, in classification it is sufficient to estimate the Bayes rule consistently, the Fisher-consistent condition is weaker than proper scoring rules. However, we show in the next section that many Fisher-consistent losses do provide estimates of the class probabilities. Thus, they can be considered as the *multicategory proper scoring rules*.

**4. Convex multicategory Fisher-consistent losses.** In this section we show that there are a number of Fisher-consistent loss functions for multicategory classification. In this work all loss functions are assumed to be non-negative. Without loss of generality, we assume $\arg\max_{c_i \in \mathcal{C}} p(y = c_i|\mathbf{x})$ is unique. We have the following sufficient condition for a differentiable convex function to be universally Fisher-consistent.

THEOREM 1. *Let $\phi(t)$ be a twice differentiable loss function. If $\phi'(0) < 0$ and $\phi''(t) > 0\ \forall t$, then $\phi$ is universally Fisher-consistent. Moreover, letting $\hat{f}$ be the solution of (3.4), then we have*

$$p(y = j|x) = \frac{1/\phi'(\hat{f}_j(x))}{\sum_{k=1}^{m} 1/\phi'(\hat{f}_k(x))}. \tag{4.1}$$

Theorem 1 immediately concludes that the two most popular smooth loss functions, namely, exponential loss and logistic regression loss (also called logit loss hereafter), are universally Fisher-consistent for multicategory classification. The inversion formula (4.1) also shows that once the margin vector is obtained, one can easily construct estimates for the conditional class probabilities. It is remarkable because we can not only do classification but also estimate the conditional class probabilities without using the likelihood approach.

The conditions in Theorem 1 can be further relaxed without weakening the conclusion. Supposing $\phi$ satisfies the conditions in Theorem 1, we can consider the linearized version of $\phi$. Define the set $A$ as given in the proof of Theorem 1 (see Section 6) and let $t_1 = \inf A$. If $A$ is empty, we let $t_1 = \infty$. Choosing a $t_2 < 0$, then we define a new convex loss as follows:

$$\zeta(t) = \begin{cases} \phi'(t_2)(t - t_2) + \phi(t_2), & \text{if } t \leq t_2, \\ \phi(t), & \text{if } t_2 < t < t_1, \\ \phi(t_1), & \text{if } t_1 \leq t. \end{cases}$$

As a modified version of $\phi$, $\zeta$ is a decreasing convex function and approaches infinity linearly. We show that $\zeta$ is also universally Fisher-consistent.

THEOREM 2. *$\zeta(t)$ is universally Fisher-consistent and (4.1) holds for $\zeta$.*

Theorem 2 covers the squared hinge loss and the modified Huber loss. Thus, Theorems 1 and 2 conclude that the popular smooth loss functions used in binary classification are universally Fisher-consistent for multicategory classification. In the reminder of this section we closely examine these loss functions.

4.1. *Exponential loss.* We consider the case $\phi_1(t) = e^{-t}$, $\phi_1'(t) = -e^{-t}$ and $\phi_1''(t) = e^{-t}$. By Theorem 1, we know that the exponential loss is universally Fisher-consistent. In addition, the inversion formula (4.1) in Theorem 1 tells us that

$$p_j = \frac{e^{\hat{f}_j}}{\sum_{k=1}^{m} e^{\hat{f}_k}}.$$

To express $\hat{f}$ by $p$, we write

$$\hat{f}_j = \log(p_j) + \log\left(\sum_{k=1}^m e^{\hat{f}_k}\right).$$

Since $\sum_{j=1}^m \hat{f}_j = 0$, we conclude that

$$0 = \sum_{j=1}^m \log(p_j) + m\log\left(\sum_{k=1}^m e^{\hat{f}_k}\right),$$

or equivalently,

$$\hat{f}_j = \log(p_j) - \frac{1}{m}\sum_{k=1}^m \log(p_k).$$

Thus, the exponential loss derives exactly the same estimates by the multinomial deviance function.

4.2. *Logit loss.* The logit loss function is $\phi_2(t) = \log(1 + e^{-t})$, which is essentially the negative binomial deviance. We compute $\phi'_2(t) = \frac{-1}{1+e^t}$ and $\phi''_2(t) = \frac{e^t}{(1+e^t)^2}$. Then Theorem 1 says that the logit loss is universally Fisher-consistent. By the inversion formula (4.1), we also obtain

$$p_j = \frac{1 + e^{\hat{f}_j}}{\sum_{k=1}^m (1 + e^{\hat{f}_k})}.$$

To better appreciate formula (4.1), let us try to express the margin vector in terms of the class probabilities. Let $\lambda^* = \sum_{k=1}^m (1 + e^{\hat{f}_k})$. Then we have

$$\hat{f}_j = \log(-1 + p_j\lambda^*).$$

Note that $\sum_j^p \hat{f}_j = 0$, thus, $\lambda^*$ is the root of equation

$$\sum_{j=1}^m \log(-1 + p_j\lambda) = 0.$$

When $m = 2$, it is not hard to check that $\lambda^* = p_1 p_2$. Hence, $\hat{f}_1 = \log(\frac{p_1}{p_2})$ and $\hat{f}_2 = \log(\frac{p_2}{p_1})$, which are the familiar results for binary classification. When $m > 2$, $\hat{f}$ depends on $p$ in a much more complex way. But $p$ is always easily computed from the margin vector $\hat{f}$.

The logit loss is quite unique, for it is essentially the negative (conditional) log-likelihood in the binary classification problem. In the multicategory problem, from the likelihood point of view, the multinomial likelihood should be used, not the logit loss. From the viewpoint of the Fisher-consistent loss, the logit loss is also appropriate for the multicategory classification problem, because it is universally Fisher-consistent. We later demonstrate the usefulness of the logit loss in multicategory classification by deriving a multicategory logit boosting algorithm.

4.3. *Least squares loss, Squared hinge loss and modified Huber loss.* The least squares loss is $\phi_3(t) = (1-t)^2$. We compute $\phi_3'(t) = 2(t-1)$ and $\phi_3''(t) = 2$. $\phi'(0) = -2$, hence, by Theorem 1, the least squares loss is universally Fisher-consistent. Moreover, the inversion formula (4.1) shows that

$$p_j = \frac{1/(1-\hat{f}_j)}{\sum_{k=1}^m 1/(1-\hat{f}_k)}.$$

We observe that $\hat{f}_j = 1 - (p_j \lambda_*)^{-1}$, where $\lambda_* = \sum_{k=1}^m 1/(1-\hat{f}_k)$. $\sum_{j=1}^p \hat{f}_j = 0$ implies that $\lambda_*$ is the root of equation $\sum_{j=1}^m (1 - (\lambda p_j)^{-1}) = 0$. We solve $\lambda_* = \frac{1}{m}(\sum_{j=1}^m 1/p_j)$. Thus,

$$\hat{f}_j = 1 - \frac{1}{p_j} \cdot \left( \frac{1}{m} \sum_{k=1}^m 1/p_k \right)^{-1}.$$

When $m = 2$, we have the familiar result: $\hat{f}_1 = 2p_1 - 1$, by simply using $1/p_1 + 1/p_2 = 1/p_1 p_2$. In multicategory problems the above formula says that with the least squares loss, the margin vector is directly linked to the inverse of the conditional class probability.

We consider $\phi_4(t) = (1-t)_+^2$, where "+" means the positive part. $\phi_4$ is called the squared hinge loss. It can be seen as a linearized version of least squares loss with $t_1 = 1$ and $t_2 = -\infty$. By Theorem 2, the squared hinge loss is universally Fisher-consistent. Furthermore, it is interesting to note that the squared hinge loss shares the same population minimizer with least squares loss.

Modified Huber loss is another linearized version of least squares loss with $t_1 = 1$ and $t_2 = -1$, which is expressed as follows:

$$\phi_5(t) = \begin{cases} -4t, & \text{if } t \leq -1, \\ (t-1)^2, & \text{if } -1 < t < 1, \\ 0, & \text{if } 1 \leq t. \end{cases}$$

By Theorem 2, we know modified Huber loss is universally Fisher-consistent. The first derivative of $\phi_5$ is

$$\phi_5'(t) = \begin{cases} -4, & \text{if } t \leq -1, \\ 2(t-1), & \text{if } -1 < t < 1, \\ 0, & \text{if } 1 \leq t, \end{cases}$$

which is used to convert the margin vector to the conditional class probability.

MULTICATEGORY BOOSTING AND FISHER-CONSISTENT LOSSES 9**Algorithm 5.1** *Multicategory GentleBoost*

1. Start with $w_i = 1$, $i = 1, 2, \ldots, n$, $G_j(\mathbf{x}) = 0$, $j = 1, \ldots, m$.
2. For $k = 1$ to $M$, repeat:
   (a) For $j = 1$ to $m$, repeat:
      i. Let $z_i = -1/m + I(y_i = j)$. Compute $w_i^* = w_i z_i^2$ and re-normalize.
      ii. Fit the regression function $g_j(\mathbf{x})$ by weighted least-squares of working response $z_i^{-1}$ to $\mathbf{x}_i$ with weights $w_i^*$.
      iii. Update $G_j(\mathbf{x}) = G_j(\mathbf{x}) + g_j(\mathbf{x})$.
   (b) Compute $f_j(\mathbf{x}) = G_j(\mathbf{x}) - \frac{1}{m}\sum_{k=1}^{m} G_k(\mathbf{x})$.
   (c) Compute $w_i = \exp(-f_{y_i}(\mathbf{x}_i))$.
3. Output the classifier $\arg\max_j f_j(\mathbf{x})$.

**5. Multicategory boosting algorithms.** In this section we take advantage of the multicategory Fisher-consistent loss functions to construct multicategory classifiers that treat all classes simultaneously without reducing the multicategory problem to a sequence of binary classification problems. We follow Friedman, Hastie and Tibshirani (2000) and Friedman (2001) to view boosting as a gradient decent algorithm that minimizes the exponential loss. This view was also adopted by Bühlmann and Yu (2003) to derive $L_2$-boosting. For a nice overview of boosting, we refer the readers to Bühlamnn and Hothorn (2007). Borrowing the gradient decent idea, we show that some new multicategory boosting algorithms naturally emerge when using multicategory Fisher-consistent losses.

5.1. *GentleBoost.* Friedman, Hastie and Tibshirani (2000) proposed the binary Gentle AdaBoost algorithm to minimize the exponential loss by using regression trees as base learners. In the same spirit we can derive the *multicategory GentleBoost* algorithm, as outlined in Algorithm 5.1.

5.1.1. *Derivation of GentleBoost.* By the symmetry constraint on $f$, we consider the following representation:

$$(5.1) \qquad f_j(\mathbf{x}) = G_j(\mathbf{x}) - \frac{1}{m}\sum_{k=1}^{m} G_k(\mathbf{x}) \qquad \text{for } j = 1, \ldots, m.$$

No restriction is put on $G$. We write the empirical risk in terms of $G$:

$$(5.2) \qquad \frac{1}{n}\sum_{i=1}^{n} \exp\left(-G_{y_i}(\mathbf{x}_i) + \frac{1}{m}\sum_{k=1}^{m} G_k(\mathbf{x}_i)\right) := L(G).$$

We want to find increments on $G$ such that the empirical risk decreases most. Let $g(\mathbf{x})$ be the increments. Following the derivation of the Gentle AdaBoost algorithm in Friedman, Hastie and Tibshirani (2000), we consider

---

**Algorithm 5.2** *AdaBoost.ML*

1. Start with $f_j(\mathbf{x}) = 0$, $j = 1, \ldots, m$.
2. For $k = 1$ to $M$:
   (a) Compute weights $w_i = \frac{1}{1+\exp(f_{y_i}(\mathbf{x}_i))}$ and re-normalize.
   (b) Fit a $m$-class classifier $T_k(\mathbf{x})$ to the training data using weights $w_i$. Define
   $$g_j(\mathbf{x}) = \begin{cases} \sqrt{\dfrac{m-1}{m}}, & \text{if } T_k(\mathbf{x}) = j, \\ -\sqrt{\dfrac{1}{m(m-1)}}, & \text{if } T_k(\mathbf{x}) \neq j. \end{cases}$$
   (c) Compute $\hat{\gamma}_k = \arg\min_\gamma \frac{1}{n}\sum_{i=1}^n \log(1+\exp(-f_{y_i}(\mathbf{x}_i) - \gamma g_{y_i}(\mathbf{x}_i)))$.
   (d) Update $f(\mathbf{x}) \leftarrow f(\mathbf{x}) + \hat{\gamma}_k g(\mathbf{x})$.
3. Output the classifier $\arg\max_j f_j(\mathbf{x})$.

---

the expansion of (5.2) to the second order and use a diagonal approximation to the Hessian, then we obtain

$$L(G+g) \approx L(G) - \frac{1}{n}\sum_{i=1}^n \left(\sum_{k=1}^m g_k(\mathbf{x}_i) z_{ik} \exp(-f_{y_i}(\mathbf{x}_i))\right)$$
$$+ \frac{1}{n}\sum_{i=1}^n \frac{1}{2}\left(\sum_{k=1}^m g_k^2 z_{ik}^2(\mathbf{x}_i) \exp(-f_{y_i}(\mathbf{x}_i))\right),$$

where $z_{ik} = -1/m + I(y_i = k)$. For each $j$, we seek $g_j(\mathbf{x})$ that minimizes

$$-\sum_{i=1}^n g_j(\mathbf{x}_i) z_{ij} \exp(-f_{y_i}(\mathbf{x}_i)) + \sum_{i=1}^n \frac{1}{2} g_j^2(\mathbf{x}_i) z_{ij}^2 \exp(-f_{y_i}(\mathbf{x}_i)).$$

A straightforward solution is to fit the regression function $g_j(\mathbf{x})$ by weighted least-squares of $z_{ij}^{-1}$ to $\mathbf{x}_i$ with weights $z_{ij}^2 \exp(-f_{y_i}(\mathbf{x}_i))$. Then $f$ is updated accordingly by (5.1). In the implementation of the multicategory Gentle-Boost algorithm we use regression trees to fit $g_j(\mathbf{x})$.

5.2. *AdaBoost.ML.* We propose a new logit boosting algorithm (Algorithm 5.2) by minimizing the binary logit risk. Similar to AdaBoost, the new logit boosting algorithm aggregates the multicategory decision tree, thus, we call it *AdaBoost.ML*.

5.2.1. *Derivation of AdaBoost.ML.* We use the gradient decent algorithm to find $\hat{f}(\mathbf{x})$ in the space of margin vectors to minimize

$$\text{EER}_n(f) = \frac{1}{n}\sum_{i=1}^n \log(1+\exp(-f_{y_i}(\mathbf{x}_i))).$$

Supposing $f(\mathbf{x})$ is the current fit, the negative gradient of the empirical logit risk is $(\frac{1}{n} \cdot \frac{1}{1+\exp(f_{y_i}(\mathbf{x}_i))})_{i=1,\ldots,n}$. After normalization, we can take the negative gradient as $(w_i)_{i=1,\ldots,n}$, the weights in 2(a).

Second, we find the optimal incremental direction $g(\mathbf{x})$, which is a functional in the margin-vector space and best approximates the negative gradient direction. Thus, we need to solve the following optimization problem:

$$(5.3) \quad \arg\max \sum_{i}^{n} w_i g_{y_i}(\mathbf{x}_i) \quad \text{subject to} \quad \sum_{j=1}^{m} g_j = 0 \quad \text{and} \quad \sum_{j=1}^{m} g_j^2 = 1.$$

On the other hand, we want to aggregate multicategory classifiers, thus, the increment function $g(\mathbf{x})$ should be induced by a $m$-class classifier $T(\mathbf{x})$. Consider a simple mapping from $T$ to $g$

$$g_j(\mathbf{x}) = \begin{cases} a, & \text{if } j = T(\mathbf{x}), \\ -b, & \text{if } j \neq T(\mathbf{x}), \end{cases}$$

where $a > 0$ and $b > 0$. The motivation of using the above rule comes from the proxy interpretation of the margin. The classifier $T$ predicts that class $T(\mathbf{x})$ has the highest conditional class probability at $\mathbf{x}$. Thus, we increase the margin of class $T(\mathbf{x})$ by $a$ and decrease the margin of other classes by $b$. The margin of the predicted class relatively gains $(a+b)$ against other less favorable classes. We decrease the margins of the less favorable classes simply to satisfy the sun-to-zero constraint. By the constraints in (5.3), we have

$$0 = \sum_{j=1}^{n} g_j = a - (m-1)b \quad \text{and} \quad 1 = \sum_{j=1}^{n} g_j = a^2 + (m-1)b^2.$$

Thus, $a = \sqrt{1 - 1/m}$ and $b = 1/\sqrt{m(m-1)}$. Observe that

$$\sum_{i=1}^{n} w_i g_{y_i}(\mathbf{x}_i) = \left(\sum_{i \in CC} w_i\right)\sqrt{1 - 1/m} - \left(\sum_{i \in NC} w_i\right) 1/\sqrt{m(m-1)},$$

where

$$CC = \{i : y_i = T(\mathbf{x}_i)\} \quad \text{and} \quad NC = \{i : y_i \neq T(\mathbf{x}_i)\}.$$

Thus, we need to find a classifier $T$ to maximize $\sum_{i \in CC}^{n} w_i$, which amounts to fitting a classifier $T(\mathbf{x})$ to the training data using weights $w_i$. The fitted classifier $T(\mathbf{x})$ induces the incremental function $\hat{g}(\mathbf{x})$.

Then for a given incremental direction $g(\mathbf{x})$, in 2(d) we compute the step length by solving

$$\hat{\gamma} = \arg\min_{\gamma} \frac{1}{n} \sum_{i=1}^{n} \log(1 + \exp(-f_{y_i}(\mathbf{x}_i) - \gamma g_{y_i}(\mathbf{x}_i))).$$

The updated fit is $f(\mathbf{x}) + \hat{\gamma}\hat{g}(\mathbf{x})$. The above procedure is repeated $M$ times.

TABLE 1
*Data sets used in the experiments*

| Data | No. Train | No. Test | Inputs | Classes | CART error |
|---|---|---|---|---|---|
| Waveform | 300 | 5000 | 21 | 3 | 31.6% |
| Vowel | 528 | 462 | 10 | 11 | 54.1% |
| Optdigits | 3823 | 1797 | 64 | 10 | 16.6% |
| Image segmentation | 210 | 2100 | 19 | 7 | 9.8% |
| Pendigits | 7494 | 3498 | 16 | 10 | 8.32% |

5.3. *Some experiments with real-world data.* Here we show the results of comparing the three multicategory boosting algorithms, AdaBoost.MH, GentleBoost and AdaBoost.ML, on several benchmark data sets obtained from the UCI machine learning repository [Newman, Hettich and Merz (1998)]. The number of boosting steps was 200 in all algorithms and examples. For reference, we also fit a single decision tree on each data set. The purpose of the experiments is to demonstrate the validity of our new multicategory boosting algorithms.

We fixed the tree size in four algorithms. The decision stumps are commonly used as base learners in AdaBoost, and hence in AdaBoost.MH. In AdaBoost.ML, we require each base learner $T_k$ to be a weak classifier for the $m$-class problem (the accuracy of $T_k$ is better than $1/m$). In the binary classification case, two-node trees are generally sufficient for that purpose. Similarly, we suggest using classification trees with (at least) $m$ terminal nodes in $m$-class problems. GentleBoost combines regression trees. The chosen value for the number of terminal nodes ($J$) should reflect the level of dominant interactions in $\hat{f}(\mathbf{x})$ [Hastie, Tibshirani and Friedman (2001)]. $J = 2$ is often inadequate, and $J \geq 10$ is also very unlikely. Following the suggestion in Hastie, Tibshirani and Friedman (2001), we used 8-node regression trees in GentleBoost.

Table 1 summarizes these data sets and the test error rates using a single decision tree. Table 2 shows the test error rates. Figure 1 displays the test error curves of the four algorithms on `waveform` and `vowel`. The test-error curves of GentleBoost and AdaBoost.ML show the characteristic pattern of a boosting procedure: the test error steadily decreases as the boosting iterations proceed and then stays (almost) flat. These experiments clearly show that the new algorithms work well and have very competitive performances as AdaBoost.MH. GentleBoost seems to perform slightly better than AdaBoost.MH.

We do not intend to argue that the new algorithms always outperform AdaBoost.MH. In fact, AdaBoost.MH is asymptotical optimal [Zhang (2004a)], thus, it is almost impossible to have a competitor that can always outperform

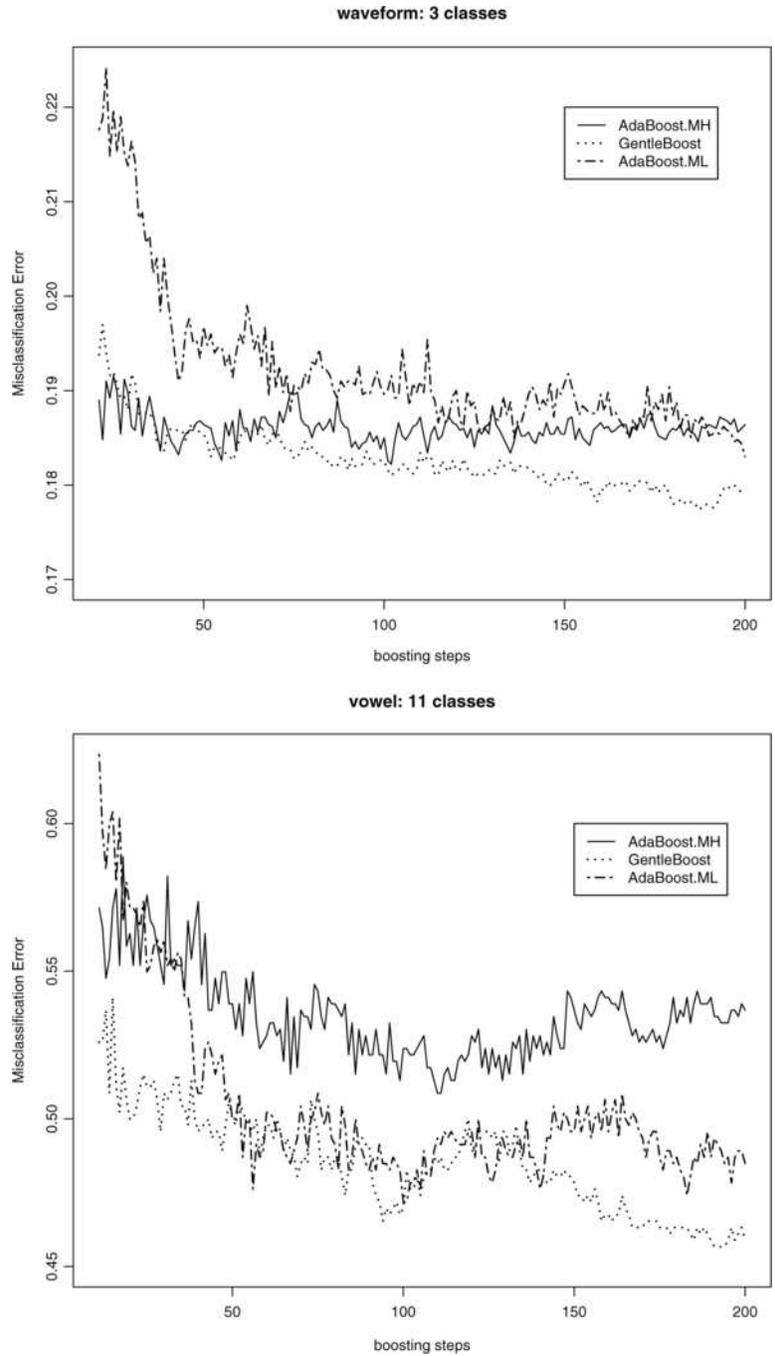

Fig. 1. Waveform *and* vowel *data: test error rate as a function of boosting steps. To better show the differences among the three algorithms, we start the plots from step 21 for* waveform *data and step 11 for* vowel *data.*

AdaBoost.MH. We are satisfied with the fact that our new multicategory boosting algorithms can do as well as AdaBoost.MH and sometimes perform slightly better than AdaBoost.MH. The working algorithms demonstrate the usefulness of the multicategory Fisher-consistent loss functions.

**6. Conclusion.** In this paper we have proposed the multicategory Fisher-consistent condition and characterized a family of convex losses that are universally Fisher-consistent for multicategory classification. To show the usefulness of the multicategory Fisher-consistent loss functions, we have also derived some new multicategory boosting algorithms by minimizing the empirical loss. These new algorithms have been empirically tested on several benchmark data sets. Fisher-consistency is the first step to establish the Bayes risk consistency of the multicategory boosting algorithms [Lin (2004), Zhang (2004a)]. It is interesting to prove multicategory GentleBoost and AdaBoost.ML converge to the Bayes classifier in terms of classification error. In future work we will follow Koltchinskii and Panchenko (2002), Blanchard, Lugosi and Vayatis (2004), Lugosi and Vayatis (2004), Bühlmann and Yu (2003) and Zhang (2004b) to study the convergence rate of the proposed multicategory boosting algorithms.

## APPENDIX: PROOFS

PROOF OF THEOREM 1. By definition of the Fisher-consistent loss, we need to show that (3.4) has a unique solution and the condition (3.5) is satisfied. Using the Lagrangian multiplier method, we define

$$L(f) = \phi(f_1)p_1 + \cdots + \phi(f_m)p_m + \lambda(f_1 + \cdots + f_m).$$

TABLE 2
*Comparing GentleBoost and AdaBoost.ML with AdaBoost.MH. Inside (·) are the standard errors of the test error rates*

|  | **AdaBoost.MH** | **AdaBoost.ML** | **GentleBoost** |
|---|---|---|---|
| Waveform | 18.22% | 18.30% | 17.74% |
|  | (0.55%) | (0.55%) | (0.54%) |
| Vowel | 50.87% | 47.18% | 45.67% |
|  | (7.07%) | (7.06%) | (7.04%) |
| Opdigits | 5.18% | 5.40% | 5.01% |
|  | (0.52%) | (0.53%) | (0.51%) |
| Image segmentation | 5.29% | 5.42% | 5.38% |
|  | (0.48%) | (0.49%) | (0.49%) |
| Pendigits | 5.86% | 4.09% | 3.69% |
|  | (0.40%) | (0.33%) | (0.32%) |

Then we have

(6.1) $$\frac{\partial L(f)}{\partial f_j} = \phi'(f_j)p_j + \lambda = 0, \qquad j = 1, \ldots, m.$$

$\phi''(t) > 0\ \forall t$, hence, $\phi'$ has an inverse function, denoted by $\psi$. Equation (6.1) gives $f_j = \psi(-\frac{\lambda}{p_j})$. By the constraint on $f$, we have

(6.2) $$\sum_{j=1}^{m} \psi\left(-\frac{\lambda}{p_j}\right) = 0.$$

$\phi'$ is a strict monotonously increasing function, so is $\psi$. Thus, the left-hand side (LHS) of (6.2) is a decreasing function of $\lambda$. It suffices to show that equation (6.2) has a root $\lambda^*$, which is the unique root. Then it is easy to see that $\hat{f}_j = \psi(-\frac{\lambda^*}{p_j})$ is the unique minimizer of (3.4), for the Hessian matrix of $L(f)$ is a diagonal matrix and the $j$th diagonal element is $\frac{\partial^2 L(f)}{\partial f_j^2} = \phi''(f_j) > 0$. Note that when $\lambda = -\phi'(0) > 0$, we have $\frac{\lambda}{p_j} > -\phi'(0)$, then $\psi(-\frac{\lambda}{p_j}) < \psi(\phi'(0)) = 0$. So the LHS of (6.2) is negative when $\lambda = -\phi'(0) > 0$. On the other hand, let us define $A = \{a : \phi'(a) = 0\}$. If $A$ is an empty set, then $\phi'(t) \to 0-$ as $t \to \infty$ (since $\phi$ is a convex loss). If $A$ is not empty, denote $a^* = \inf A$. By the fact $\phi'(0) < 0$, we conclude $a^* > 0$. Hence, $\phi'(t) \to 0-$ as $t \to a^*-$. In both cases, we see that $\exists$ a small enough $\lambda_0 > 0$ such that $\psi(-\frac{\lambda_0}{p_j}) > 0$ for all $j$. So the LHS of (6.2) is positive when $\lambda = \lambda_0 > 0$. Therefore, there must be a positive $\lambda^* \in (\lambda_0, -\phi'(0))$ such that equation (6.2) holds.

For (3.5), let $p_1 > p_j\ \forall j \neq 1$, then $-\frac{\lambda^*}{p_1} > -\frac{\lambda^*}{p_j}\ \forall j \neq 1$, so $\hat{f}_1 > \hat{f}_j\ \forall j \neq 1$. Using (6.1), we get $p_j = -\frac{\lambda^*}{\phi'(\hat{f}_j)}$. $\sum_{j=1}^{m} p_j = 1$ requires

$$\sum_{j=1}^{m}\left(-\frac{\lambda^*}{\phi'(\hat{f}_j)}\right) = 1.$$

So it follows that $\lambda^* = -(\sum_{j=1}^{m} 1/\phi'(\hat{f}_j))^{-1}$. Then (4.1) is obtained. $\square$

PROOF OF THEOREM 2. First, by the convexity of $\zeta$ and the fact $\zeta \geq \phi(t_1)$, we know that the minimizer of (3.4) always exists. We only need to show the uniqueness of the solution and (3.5). Without loss of generality, let $p_1 > p_2 \geq p_3 \geq \cdots \geq p_{m-1} > p_m$. Suppose $\hat{f}$ is a minimizer. Substituting $f_m = -(\sum_{j=1}^{m-1} f_j)$ into (3.2), we have

(6.3) $$\zeta(p, f) = \sum_{j=1}^{m} \zeta(f_j)p_j = \sum_{j=1}^{m-1} \zeta(f_j)p_j + \zeta\left(-\left(\sum_{j=1}^{m-1} f_j\right)\right)p_m.$$

Differentiating (6.3) yields

$$\zeta'(\hat{f}_j)p_j - \zeta'(\hat{f}_m)p_m = 0, \qquad j = 1, 2, \ldots, m-1,$$

or equivalently,

(6.4) $\zeta'(\hat{f}_j)p_j = -\lambda, \qquad j = 1, 2, \ldots, m \quad$ for some $\lambda$.

There is one and only one such $\lambda$ satisfying (6.4). Otherwise, let $\lambda_1 > \lambda_2$ and $\hat{f}(\lambda_1)$, $\hat{f}(\lambda_2)$ such that

$$\zeta'(\hat{f}_j(\lambda_1))p_j = -\lambda_1, \qquad \zeta'(\hat{f}_j(\lambda_2))p_j = -\lambda_2 \quad \forall j.$$

Then we see that $\zeta'(\hat{f}_j(\lambda_1)) < \zeta'(\hat{f}_j(\lambda_2))$, so $\hat{f}_j(\lambda_1) < \hat{f}_j(\lambda_2)$ for all $j$. This is clearly a contradiction to the fact that both $\hat{f}(\lambda_1)$ and $\hat{f}(\lambda_2)$ satisfy the constraint $\sum_{j=1}^m f_j = 0$.

Observe that if $0 > \zeta'(t) > \phi'(t_2)$, $\zeta'$ has an inverse denoted as $\psi$. $\exists$ a small enough $\lambda_0$: $-\phi'(t_2)p_m > \lambda_0 > 0$ such that $\psi(-\frac{\lambda_0}{p_j})$ exists and $\psi(-\frac{\lambda_0}{p_j}) > 0$ for all $j$. Thus, the $\lambda$ in (6.4) must be larger than $\lambda_0$. Otherwise $\hat{f}_j > \psi(-\frac{\lambda_0}{p_j}) > 0$ for all $j$, which clearly contradicts $\sum_{j=1}^m f_j = 0$. Furthermore, $\zeta'(t) \geq \phi'(t_2)$ for all $t$, so $\lambda \leq -\phi'(t_2)p_m$. Then let us consider the following two situations:

Case 1. $\lambda \in (\lambda_0, -\phi'(t_2)p_m)$. Then $\psi(-\frac{\lambda}{p_j})$ exists $\forall j$, and $\hat{f}_j = \psi(-\frac{\lambda}{p_j})$ is the unique minimizer.

Case 2. $\lambda = -\phi'(t_2)p_m$. Similarly, for $j \leq (m-1)$, $\psi(-\frac{\lambda}{p_j})$ exists, and $\hat{f}_j = \psi(-\frac{\lambda}{p_j})$. Then $\hat{f}_m = -\sum_{j=1}^{m-1} \psi(-\frac{\lambda}{p_j})$.

Therefore, we prove the uniqueness of the minimizer $\hat{f}$. For (3.5), note that $\zeta'(\hat{f}_1) = -\frac{\lambda}{p_1} > -\frac{\lambda}{p_j} = \zeta'(\hat{f}_j)$ for $j \geq 2$, hence, we must have $\hat{f}_1 > \hat{f}_j \ \forall j$, due to the convexity of $\zeta$. The formula (4.1) follows (6.4) and can be derived using the same arguments as in the proof of Theorem 1. $\square$

**Acknowledgments.** We thank an anonymous referee for providing us helpful comments and suggestions that greatly improved the paper.

Hui Zou
School of Statistics
University of Minnesota
Minneapolis, Minnesota 55455
USA
E-mail: hzou@stat.umn.edu

Ji Zhu
Department of Statistics
University of Michigan
Ann Arbor, Michigan 48109
USA
E-mail: jizhu@umich.edu

Trevor Hastie
Department of Statistics
Stanford University
Stanford, California 94305
USA
E-mail: hastie@stanford.edu